\documentclass[smallextended]{svjour3}       
\usepackage{amsmath,amssymb}

\usepackage{changepage}

\usepackage[utf8x]{inputenc}

\usepackage{textcomp,marvosym}

\usepackage{cite}

\usepackage{nameref,hyperref}

\usepackage[right]{lineno}

\usepackage[aboveskip=1pt,labelfont=bf,labelsep=period,justification=raggedright,singlelinecheck=off]{caption}

\makeatletter
\renewcommand{\@biblabel}[1]{\quad#1.}
\makeatother

\usepackage{lastpage,fancyhdr,graphicx}
\usepackage{epstopdf}

\usepackage{amsmath,amsfonts,amssymb}
\renewcommand{\qed}{\hfill $\Box$}

\newcommand{\field}[1]{\mathbb{#1}}
\newcommand {\R}{\field{R} }
\newcommand {\N}{ {\field{N}} }

\newcommand {\ol}{\overline }
\newcommand {\e}{ {\mathbf e} }

\newtheorem{propx}{Proposition}
\newtheorem{ddef}[propx]{Definition}
\newtheorem{rem}[propx]{Remark}
\newtheorem{prob}[propx]{Problem}
\newtheorem{lem}[propx]{Lemma}
\newtheorem{corr}[propx]{Corollary}

\usepackage{color}
\usepackage{colordvi}
\definecolor{magenta}{rgb}{.5,0,.5}
\definecolor{black}{rgb}{1.0,1.0,1.0}
\definecolor{magenta}{rgb}{.1,0,.3}
\definecolor{gruen}{rgb}{0.2,0.5,.5}
\definecolor{light}{rgb}{ 0.992, 0.961,  0.902}
\definecolor{Tan}{rgb}{ 0.992, 0.9,  0.902}

\newcommand{\komment}[1]{{}}

\usepackage{mathptmx}      

\begin{document}
	
	\title{Homogeneous and Heterogeneous Response of Quorum-Sensing Bacteria in an Evolutionary Context}
	\titlerunning{Homogeneous and Heterogeneous Response of Quorum-Sensing Bacteria}

	\author{Hense, Burkhard \and
		McIntosh, Matthew \and
		M\"uller Johannes \and
		Schuster, Martin}

\institute{Hense, Burkhard (deceased) \at Institute of Computational Biology, Helmholtz Zentrum M\"unchen - German Research Center for Environmental Health, Ingolstädter Landstr. 1, D-85764 Neuherberg, Germany \and 
McIntosh, Matthew \at Department of Microbiology and Molecular Biology, Univ.\ of Giessen, Heinrich-Buff-Ring 26--31
		D-35392, Germany
\and
M\"uller Johannes \at  Centre for Mathematical Sciences, Technische Universit\"at M\"unchen, Boltzmannstr. 3, D-85747 Garching/Munich, Germany,
johannes.mueller@mytum.de\\
Institute of Computational Biology, Helmholtz Zentrum M\"unchen - German Research Center for Environmental Health, Ingolstädter Landstr. 1, D-85764 Neuherberg, Germany \and 
Schuster, Martin \at  Department of Microbiology, Oregon State Univ., Nash Hall 226, 
		Corvallis, OR  97331
}

	\date{Received: date / Accepted: date}

	\maketitle

\begin{abstract}
	To explain the stability of cooperation is a central task of evolutionary theory. We investigate this question in the case of quorum sensing (QS) bacteria, which regulate cooperative traits in response to population density. Cooperation is modeled by the prisoner’s dilemma, where individuals produce a costly public good (PG) that equally benefits all members of a community divided into multiple, distinct patches (multilevel selection). Cost and benefit are non-linear functions of the PG production. The analysis of evolutionary stability yields an optimization problem for the expression of PG in dependency on the number of QS individuals within a colony. We find that the optimal total PG production of the QS population mainly depends on the shape of the benefit. A graded and a switch-like response is possible, in accordance with earlier results. Interestingly, at the level of the individual cell, the QS response is determined by the shape of the costs. All QS individuals respond either homogeneously if cost are a convex function of the PG production rate, or they respond heterogeneously with distinct ON/OFF responses if the costs are concave. The latter finding is consistent with recent experimental findings, and contradicts the usual interpretation of QS as a mechanism to establish a uniform, synchronized response of a bacterial population. 
\end{abstract}

\komment{
	\section*{Author summary}
	Even genetically identical bacteria do not behave identically under uniform environmental conditions. Isogenic bacteria are capable of forming complex societies, in which different individuals perform different tasks.  This behavior, known as division of labor, leads to a higher fitness. In the present work we focus on heterogeneity in quorum sensing (QS). QS is a 
	relatively simple mechanism widely distributed among microorganisms enabling them to sense their own population size and adopt behavioral responses that are appropriate to the population size. Often, cooperative traits require the coordinated action of a large number of individuals. Recent experimental findings indicate that not all QS individuals take part in such a synchronized action. These findings contradict our understanding of QS. Based on mathematical modeling, we propose a solution to this apparent paradox, by investigating costs and benefit of cooperative traits in the light of evolution. Our analysis indicates that primarily the shape of the costs is decisive for a homogeneous or heterogeneous response of the population.
}






\section{Introduction}

The evolutionary stability of bacterial cooperation is still not fully understood~\cite{Damore2012}. Kin selection (Hamilton's rule~\cite{Hamilton1963}) and multilevel selection~\cite{price1972,Okasha2006} are standard explanations for the appearance and stability of cooperation across all domains of life~\cite{West2003,Doebeli2004,West2007a,West2006}. Multilevel evolution takes into account that individuals do not live in a homogeneous population, but in patches or colonies, which leads to hierarchical models~\cite{Okasha2006}. However, bacteria do not act blindly,  but have means to sense aspects of their social and physical environment.  In response to this information, they show plasticity and adapt their strategy~\cite{Popat2015}: In many cases, cooperation or defection is not only determined by the genotype, but also by phenotypic plasticity. The most prominent system that controls cooperation is quorum sensing (QS): Bacteria produce and respond to a relatively cheap signaling molecule, allowing inference about the number of other QS bacteria in close proximity, or about the diffusivity of the environment~\cite{Redfield2002}. Such a regulatory system enables the population to delay a costly cooperative activity, such as the release of public goods (PG), until the population size is sufficiently large to ensure effective change of the environment~\cite{Ng2009, Pai2012}.
\par\medskip

Intuitive examples of PGs are secreted antibiotics or virulence factors, which will only effectively kill competitor species or destroy host cells if a certain minimum concentration is exceeded~\cite{Papagianni2006}. QS can also ensure efficient use of a secreted enzyme~\cite{Darch2012, Cornforth2012a}: Secretion into a large volume of surrounding medium by an isolated bacterial cell will strongly dilute the enzyme and its products by diffusion. Secretion will be efficient only if many bacteria contribute in a synchronous manner. \par\medskip

If QS is considered in the framework of multilevel selection, it can be interpreted as a way to infer information about kinship, and in this to connect kin selection and kin recognition with multilevel selection~\cite{Czaran2009,Rumbaugh2012,Schluter2016}. In many cases, cooperation under control of QS is able to invade other bacterial communities more effectively, or to better resist invasion by others~\cite{Czaran2009,Melke2010,Allen2016}. For example, QS individuals are more efficient than constitutive cooperators in controlling cheater mutants that exploit cooperation~\cite{Czaran2009}. QS individuals will stop to cooperate in patches with a high frequency of defectors (kin recognition), and in this, the tragedy of the commons can be moderated. Combinations of  methods from population genetics and the concept of multilevel selection seem to indicate that cheating mutants are not able to persist as a result of frequency-dependent selection, but are repeatedly created by spontaneous mutations and are subsequently eliminated. The observed mutants are the result of this birth-death process~\cite{VanDyken2012b}. 
\par\medskip

Most work is done for homogeneous populations. In this case, the invasion analysis leads to the optimization problem to use resources in an optimal way (see below and~\cite{Heilmann2015a}). It has been argued that either a gradual increase of cooperativeness or a synchronized all-or-nothing strategy can be the best choice for the population, depending on the phenotype under control~\cite{Heilmann2015a}. Observed heterogeneity has been attributed to spatial effects or general environmental heterogeneity~\cite{Hense2012}. That is, the idea here is that the environment that QS individuals sense is subject to variability. However, one of the central assumptions of QS is that it serves to synchronize populations in their response, even if the environment is heterogeneous. Only the quorum matters~\cite{Hense2007}. Differences in the environment of single population members are thought to be overwritten by the QS signal~\cite{Melke2010}. According to this explanation, heterogeneity in bacteria should not play a major role. \par\medskip

Recent experimental results show that a heterogeneous phenotype in QS populations is surprisingly common (see e.g., ~\cite{Anetzberger2009, Anetzberger2012, Lopez2009c, Lopez2010a, Grote2014, Grote2015, Wang2014, Schlueter2015} and literature cited therein). These results call for a theoretical explanation. We focus on these experimental findings and extend the multilevel selection approach to explain this behavior. We find that it is central to acknowledging the nonlinear dependency of costs and benefit on PG production. In the first part of the paper, we introduce the basic model and perform an evolutionary multilevel analysis in the case of a homogeneously  responding population and linear cost/benefit characteristics. We were able to recover expected outcomes and also to extend them. In the second part, we augment the model to allow for nonlinear dependency of cost/benefit on the PG production in a similar way as in ref.~\cite{Heilmann2015a}. Furthermore, we allow for heterogeneity, that is, for different responses of individuals. It is possible to characterize the effect of evolutionary forces on the behavior of single bacteria. Of course, not in all settings is a heterogeneous response optimal; sometimes a homogeneous response of the population is more efficient. Our model allows us to differentiate between causes that lead to a homogeneous and non-homogeneous response, respectively.\par\medskip

On the colony/population level, a gradual response as well as a sudden increase in the PG production can be observed, if the number/fraction of QS individuals in a population is increased. We find that mainly the shape of the benefit determines which of the two possibilities is optimal.\\
In contrast to the original interpretation of QS as a mechanism that ensures a homogeneous response of bacterial communities, our model strongly predicts that QS systems controlling a cooperative phenotype are expected to respond often in a heterogeneous manner if the costs depend in a concave way on the PG production.  This conclusion is based on the following considerations: as the PG is available to all cells within a colony, the fitness differences of bacteria within this colony are not affected by the amount of PG. In contrast, the costs are different for bacteria with different strategies. Therefore the costs for PG production are more important for the single-cell level than the benefit. The second observation is that the nature of the costs determines the optimal QS strategy: A homogenous response is favored if the costs increase in a convex way as a function of the production rate (to produce PG at twice the rate is less than twice as costly). A heterogenous response is favored, in contrast, if the costs increase in a concave way. We discuss plausible biological scenarios in which the costs of PG production are either concave or convex.

\section{Mathematical Model}
We performed an invasion analysis of defectors (D),  
constitutive cooperators (C) and quorum sensing 
bacteria (QS). Cooperation is, as usual for conceptual models, 
described by the prisoners dilemma~\cite{Hofbauer1998}. We do not
consider pairwise interactions, but assume that 
a cooperator produces PG continuously (at some costs) that is equally shared  
among all individuals within the community.  
The population is organized into distinct colonies, 
which leads to multilevel evolution. This setting is 
well suited for the investigation of cooperation~\cite{Okasha2006}, 
in particular in case of bacterial cooperation~\cite{McGinty2013}. 
 \\
We start with the description of a single bacterial colony consisting of two types (an abundant resident and a rare invader) in section~\ref{oneColonyOnly}. In particular, we define the fitness functions in this context. Afterwards, in section~\ref{manyColonies}, we turn to the investigation of many colonies in parallel, basically using a variant of the time-continuous Price  equation~\cite{DayGandon2006}. In section~\ref{InvasionHetero} we extend the standard prisoners dilemma allowing for nonlinear dependencies of costs/benefit on PG production and for heterogeneity in the response of the population on the level of individuals, in a similar manner as in threshold games~\cite{Deng2011}. The most important functions and parameters of the model used are listed in Table~\ref{xtab}.

\begin{table}[htb]
	\begin{tabular}{l|l}
		Symbol & meaning\\
		\hline
		$N$ & initial number of individuals within a colony\\
$R$, $I$ & number of resident/invader individuals within a colony\\
		$Q$, $Q_i$ & degree of cooperativeness (of individual i)\\
		$Q^*$      & optimal degree of cooperativeness\\
		$f(Q,Q_i)$ & fitness of individual with coop.\ $Q$, 
		in a population with coop.\ $(Q_i)_{i=1,\ldots,N}$\\
		$F_c'(0), F'(0)$ & initial growth rate of invaders in one  colony/over all colonies\\
		$x$ & fraction of residents within a colony\\
		$y$ & fraction of invaders within a colony\\
		$\ol Q_R(x)$ & degree of cooperativeness among residents (homogeneous population)\\
		$\ol Q_I(y)$ & degree of cooperativeness among invaders (homogeneous population)\\
		$E_c(.), E_g(.)$ & expectation within one colony/over all colonies\\
		$c$, $b$ & costs and benefit of Prisoners Dilemma\\
		$r$ & relatedness\\
		$p$ &  initial fraction of invaders\\
		$C(Q)$ & costs for PG production at given cooperativeness\\
		$B(Q)$ & benefit of PG's produced at given cooperativeness\\
		$\Lambda_\nu(Q)$ & initial 
		slope in time of the expected invaders fraction ($\nu$ QS individuals)
	\end{tabular}
\caption{Most important symbols and functions and their interpretation.}
\label{xtab}
\end{table}

\subsection{One colony}\label{oneColonyOnly}
Let us consider a colony, founded by $N$ bacteria. 
We assume that we have a resident population with number $R_t$, 
and an invading population with number $I_t$ at time $t$. 
In the present section, 
initial conditions $R_0$ and $I_0$ are deterministically 
given and not random -- this 
assumption will be changed when we move on to the multilevel analysis. 
Each individual may always defect (D), may always cooperate (C), or may adapt its behavior according to the fraction of QS-individuals in the population (QS strategy). 
Let us number the individuals in the colony at time $t$ by 
$i\in\{1,\ldots,R_t+I_t\}$.
We assign to each individual a variable $Q_t^i$ that indicates the degree 
at which the individual cooperates: $Q_i^t=0$ for defectors or 
a non-cooperating 
QS-individual,  $Q_i^t=1$ for cooperators 
or a fully cooperating QS individual. In contrast to C and D individuals, QS individuals sense the environment and thereby acquire information about the fraction of QS individuals within the population. They can adapt the degree of cooperation accordingly. If QS-individuals produce PG at a positive but not maximal 
rate, $Q_i^t\in(0,1)$ represents 
the actual chosen production rate. This rate is normalized, s.t.\ $Q_i^t=1$ corresponds to the maximal production rate. The cell regulatory pathways of QS often incorporate a positive feedback loop. It depends on the parameters of this feedback loops if we find at the single-cell level a graduate response or an all- or nothing response~\cite{Fujimoto2013}. We assume that 
the sensing apparatus itself has negligible costs.\par\medskip

Next we define the population dynamics. All bacteria divide at a certain rate, 
but do not die. We have a time-continuous, stochastic 
birth process. The net birth rate, that takes into account costs for 
and benefit from cooperation is the fitness of an individual. 
A constitutive cooperator always has costs $c$, a QS-individual cooperating 
at degree $Q\in[0,1]$ has costs $Q\, c$. All cooperative individuals (C and QS) contribute to PG that is available to all colony members - each individual within 
the colony receives a fair share. 
The available PG per individual is simply given by  
$b\,\sum_{i=1}^{R_t+I_t}Q_i^t/(R_t+I_t)$ -- of course, there is the usual time 
scale argument behind the assumption: at a time scale that is short in comparison with reproduction, the PG concentration tends to an equilibrium. 
Let us consider a population where individual $i$ ($i=1,\ldots \nu_t = R_t+I_t$)  
cooperates at degree $Q_i$. Select one of these individuals. If this focal 
individual cooperates at degree $Q$, its fitness reads 
\begin{eqnarray}
f(Q; Q_1,\ldots,Q_{\nu_t}) &=& 1 - Q c + b\,\sum_{i=1}^{R_t+I_t}\frac{Q_i}{R_t+I_t}.
\end{eqnarray}
To guarantee a positive fitness, we assume throughout the paper that $c\in(0,1)$. In the first case a homogeneously responding population is addressed. This is the case if all individuals of a given type show the same response (degree of cooperativeness) to their social environment, such that there are functions $\ol Q_R(x)$ resp.\ $\ol Q_I(y)$ that indicate the normalized PG production rate strength of resident resp.\ invading population (where $x = R_t/(R_t+I_t)$ resp.\ $y = I_t/(R_t+I_t)$). In case of C (D) the functions are constant ($\ol Q_{R/I}(.)=1$ resp.\ $\ol Q_{R/I}(.)=0$). Only in the QS case, $\ol Q_{R/I}(.)$ indeed depends on $x$ resp.\ $y$.\par\medskip 

We assume that the resident population only consists of one type (C, D, QS), and also the invaders have all the same genotype. As we consider, at the time being, a homogeneous response, all resident (invading) individuals have the same cooperation strength $Q_i=\ol Q_R(x)$ if individual $i$ is a resident ($Q_i=\ol Q_I(y)$ if $i$ is an invader). Then, 
\begin{eqnarray}
R_t &\rightarrow& R_t+1\qquad \mbox{at rate } 1 - \ol Q_R(x) c + b\left(\,\frac{\ol Q_R(x)\, R_t+\ol Q_I(y)\,I_t}{R_t+I_t}\right)\\
I_t &\rightarrow& I_t+1\qquad \mbox{at rate }  1 - \ol Q_I(y) c + b\left(\,\frac{\ol Q_R(x)\, R_t+\ol Q_I(y)\,I_t}{R_t+I_t}\right).
\end{eqnarray}
 \\
Our main focus is the question if the invaders can spread. To be more precise, 
we want to know if the fraction of invaders becomes larger or smaller in the first, small time interval. 
Thereto we compute the expectation of this fraction for one single colony (which we denote by $E_c(.)$). 
Let $F_t^c=E_c(I_t/(R_t+I_t))$. If 
$$  F_c'(0) := \frac  d{dt} F_t^c|_{t=0} >0,$$
then the fraction of invaders increases initially. 
\begin{propx} Let $x=R_0/N$ the initial fraction of residents, and $y=1-x$, the initial fraction of invaders. With the notation introduced above, 
$$ F_c'(0) = 
c\,\frac{x\, y}{1+1/N}\,\,\left( \ol Q_R(x)-\ol Q_I(y)   \right).$$
\end{propx}
{\bf Proof: }
In a small time interval $\Delta t$, the population will either stay constant, 
or increase by one (we suppress the dependency on $\ol Q_{R/I}$ of $x$ resp.\ $y$): 
\begin{eqnarray*}
P(R_{\Delta t}=R_0,I_{\Delta t}=I_0) &=&1- R_0\left(1 - \ol Q_R c + b\left(\,\frac{\ol Q_R\, R_0+\ol Q_I\,I_0}{N}\right)\right)\Delta t\\
&&\quad-I_0\left(1 - \ol Q_I c + b\left(\,\frac{\ol Q_R\, R_0+\ol Q_I\,I_0}{N}\right)\right)\Delta t,\\
P(R_{\Delta t}=R_0+1,I_{\Delta t}=I_0)&=& R_0\left(1 - \ol Q_R c + b\left(\,\frac{\ol Q_R\, R_0+Q_i\,I_0}{N}\right)\right)\Delta t,\\
P(R_{\Delta t}=R_0,I_{\Delta t}=I_0+1)&=& I_0\left(1 - \ol Q_I c + b\left(\,\frac{\ol Q_R\, R_0+\ol Q_I\,I_0}{N}\right)\right)\Delta t,
\end{eqnarray*}
and all other transitions have a probability of ${\cal O}(\Delta t^2)$. 
Thus (note that we know $R_0$ and $I_0$, that is, $F_0$ is 
known and not a random variable)
\begin{eqnarray*}
 F_c'(0) = \lim_{\Delta t\rightarrow 0}
\frac{E(F_{\Delta t}^c)- F_0^c}{\Delta t} 
&=& c\,\frac{I_0R_0}{N(N+1)}\,\,\left( \ol Q_R-\ol Q_I   \right).
\end{eqnarray*}
\par\qed\par\medskip

The proposition indicates that an invader can spread if $\ol Q_I<\ol Q_R$, that is, if the invader is less cooperative. A central assumption here is the homogeneity of the population. Results of this type are well known in case of  Wright-Fisher and Moran models (where the total population size is constant) with frequency dependent selection~\cite{Foster1990,Imhof2006}.

\begin{rem}
The equation above is closely related to the replicator equation (which is developed for the deterministic counterpart of our stochastic process).  Standard arguments (Ethier and Kutz~\cite{Ethier2005}) indicate that the fraction $y(t)=I_t/N$ approximates a deterministic dynamical system for large $N$. From the argument above, we have ($R_t/N=(N-I_t)/N=1-y(t)$, $y_0:=I_0/N$) that $y'(0) = c\,y(0)(1-y(0)) (\,\,\ol Q_R(1-y(0))-\ol Q_I(y(0))\,\,)$. As this equation is true for any initial value $y(0)=y_0$, we obtain the ODE $y'=c\,y(1-y)(\,\,\ol Q_R(1-y)-\ol Q_I(y)\,\,)$. This is the well-known replicator equation for two species~\cite{Hofbauer1998}. In this ODE, the solution $y(t)$ always tends to a stationary point. A stationary point $y^*$ is given either by $y^*=0$, by 
$y^*=1$ or by the condition $\ol Q_R(1-y^*)=\ol Q_I(y^*)$. That is, if defectors  ($Q=0$) competes with cooperators ($Q=1$), the cooperators always loose. If cooperators ($Q=1$) compete with QS individuals $Q=\ol Q(y)\in[0,1]$, the fraction of QS individuals grows until either they all choose the strategy to fully cooperate ($Q(y)=1$), or the cooperators die out. In the same way, the fraction of QS individuals competing with a defector population shrinks until they all stop to cooperate and behave as defectors (or they die out).
\end{rem}
As a result, constitutive cooperators in a homogeneous population are never able to spread. QS individuals can only persist in a  defecting population if they all behave as defectors. Since we assume that QS individuals that do not produce PG have no costs, their fitness is that of defecting bacteria. Due to this symmetry, our model predicts that D and (non-cooperating) QS individuals can coexist, but not D and cooperating QS individuals. This result does not explain why we find cooperating QS individuals in experiments.  However, the picture will change if move on towards multilevel evolution.

\subsection{Multilevel analysis of homogeneously responding QS individuals}
\label{manyColonies}
We extend the analysis from only one colony to many colonies. For simplicity, all colonies start with the very same population size $N$. However, the initial composition of a colony is not deterministic ($I_0$ is a random variable, $R_0=N-I_0$). We introduce the expectation of a random variable over all colonies: While $E_c(I_t|I_0=i)$ denotes the expected number of invaders at time $t$ in one given colony with initial value $I_0=i$, $E_g(I_t)$ is the average number of invaders 
over all colonies, that is, 
$$ E_g(I_t) = \sum_{i=0}^N E_c(I_t|I_0=i)\,P(I_0=i).$$ 
We aim at the dynamics of the fraction of invaders over all colonies. Assume that we have $M$ colonies numbered by $j=1,\ldots, M$. In a given realization, the colony with number $j$ starts with $I_0=i_j$ invaders, where $i_j$ is selected according to some random law. The total number of invaders (summarized over all colonies) reads $\sum_{j=1}^M (I_t|I_0=i_j)$ while the total population is $\sum_{j=1}^M ((R_t+I_t)|I_0=i_j)$. We are interested in 
\begin{eqnarray} 
\frac{\sum_{j=1}^M (I_t|I_0=i_j)}{\sum_{j=1}^M ((R_t+I_t)|I_0=i_j)}
=
\frac{\frac 1 M\sum_{j=1}^M (I_t|I_0=i_j)}{\frac 1 M \sum_{j=1}^M ((R_t+I_t)|I_0=i_j)}.\label{xxxQeun}
\end{eqnarray}
We aim at $M\rightarrow\infty$; therefore, we focus on one initial condition $i_0\in\{0,\ldots, N\}$, which has probability $P(I_0=i_0)$.  In (\ref{xxxQeun}), it appears on average $M\, P(I_0=i)$ times. That is, this initial condition appears (either never a.s.\ if $P(I_0=i_0)=0$ or) very often if $M$ becomes large. The law of large numbers implies that the mean of a random variable over many realizations tends to its expectation (under wide assumptions, which are given here \cite{Bauer1996}). Thus
$$ F(t) := \lim_{ M\rightarrow\infty} \frac{\sum_{j=1}^M (I_t|I_0=i_j)}{\sum_{j=1}^M ((R_t+I_t)|I_0=i_j)} = 
 \frac{\sum_{i=1}^N E(I_t|I_0=i)P(I_0=i)}{\sum_{i=1}^N E((R_t+I_t)|I_0=i)P(I_0=i)} 
= \frac{E_g(I_t)}{E_g(R_t+I_t)} .$$
Though we consider a stochastic model, the limit $M\rightarrow\infty$ yields 
an infinite population size such that the result is deterministic. 
Therefore, the following definition  is appropriate.

\begin{ddef} Let 
$$ F'(0) = \frac d {dt}\frac{E_g(I_t)}{E_g(R_t+I_t)}\bigg|_{t=0}.$$
We say that  $I$ can invade $R$ if $F'(0)>0$.
\end{ddef}

The important difference between $F'(0)$ and $F_c'(0)$ is the 
reference population for the fraction of invaders: In $F_c'(0)$ we consider a 
single colony, in $F'(0)$ all colonies. \\
For the next proposition, recall that we still address homogeneously responding populations. 

\begin{propx} Let $x$, $y$ be random variables 
$x=R_0/N$, $y=I_0/N$, and $p$ the initial fraction of invaders, $p=E_g(y)$. Then, 
\begin{eqnarray}
 F'(0) &=& 
(1-p) E_g\left[y\,\left(-c\, \ol Q_I(y)+  b\,\left(\,\ol Q_R(x)\, x+\ol Q_I(y)\,y\right)      \right)\right]\\
&&-
p E_g\left[(1-y)\,\left(-c\, \ol Q_R(x)+b\,\left(\,\ol Q_R(x)\,x+\ol Q_I(y)\,y\right)      \right)\right]\nonumber
\end{eqnarray}
\end{propx}
{\bf Proof: }
Recall $R_0+I_0=N$ in all colonies, that is, $E_g(R_0+I_0)=N$. 
In the following computation, we partially suppress the argument of $\ol Q_I$ and $\ol Q_R$; strictly spoken, $\ol Q_I=\ol Q_I(y)$, $y=(I_t/(R_t+I_t)$ and $\ol Q_R=\ol Q_R(x)$, $x=(R_t/(R_t+I_t))$. Then, we have 
\begin{eqnarray*}
 F'(0) &=&
\frac{ \frac d {dt}E_g(I_t)}{E_g(R_t+I_t)}
- \frac{E_g(I_t)}{E_g(R_t+I_t)}\,\, \frac{ \frac d {dt}E_g(R_t+I_t)}{E_g(R_t+I_t)}
\bigg|_{t=0}\\
&=& \frac 1 N \,\,E_g \left[ I_0\left(1 - \ol Q_I c + b\,\left(\,\frac{\ol Q_R\, R_0+\ol Q_I\,I_0}{N}\right)\right)\,\,\right]\\
&&-\frac{E_g(I_0)} N\,\,\frac 1 N \,\,E_g\bigg[
R_0\left(1 - \ol Q_R c + b\,\left(\,\frac{\ol Q_R\, R_0+\ol Q_I\,I_0}{R_0+I_0}\right)\right)\\
&&\qquad\qquad\qquad\qquad+ I_0\left(1 - \ol Q_I c + b\,\left(\,\frac{\ol Q_R\, R_0+\ol Q_I\,I_0}{N}\right)\right)\,\,\bigg]\\
&=& 
(1-E_g(y))  \,\,E_g \left[ y\bigg(1 - \ol Q_I(y) c + b\,(\ol Q_R(x)\,x+\ol Q_I(y)\, y)\,\,\bigg)\,\,\right]\\
&&- E_g(y)\,\,E_g\left[
(1-y)\,\bigg(1 - \ol Q_R(x) c + b\,(\ol Q_R(x)\,x+\ol Q_I(y)\, y)\,\,\bigg)\,\,\right]
\end{eqnarray*}
The result follows with $p=E_g(y)$. 
\par\qed\par\medskip

We can use this proposition to investigate the conditions that are necessary for an invasion of D by C, an invasion of D by QS, and an invasion of C by QS, where we only allow for a homogeneous response (all residents resp.\ all invaders have the same $Q$)\\
Recall $\mbox{var}_g(y) = E_g(y^2)-(E_g(y))^2$. The next proposition reproduces the well-known Hamilton's rule. 

\begin{propx} C can invade D, if
$$ c < r\, b,\qquad r = \frac{\mbox{var}_g(y)}{E_g(y)\,(1-E_g(y))}.$$
\end{propx}
{\bf Proof: }As the resident is a defector and the invader is a constitutive cooperator, we have  $\ol Q_R=0$ and $\ol Q_I=1$. Thus, with $p=E_g(y)$, 
\begin{eqnarray*}
 F'(0) &=& 
(1-p) E_g\left[y\,\left(-c\,+  b\,y     \right)\right]
-
p E_g\left[(1-y)\,b\,y    \right]\\
&=& -c\, p\,(1-p)+b\,(E_g(y^2)-p^2) = p(1-p)\big[b\, r-c\big].
\end{eqnarray*}
\par\qed\par
\begin{figure}[htb]
\begin{center}
\includegraphics[width=12cm]{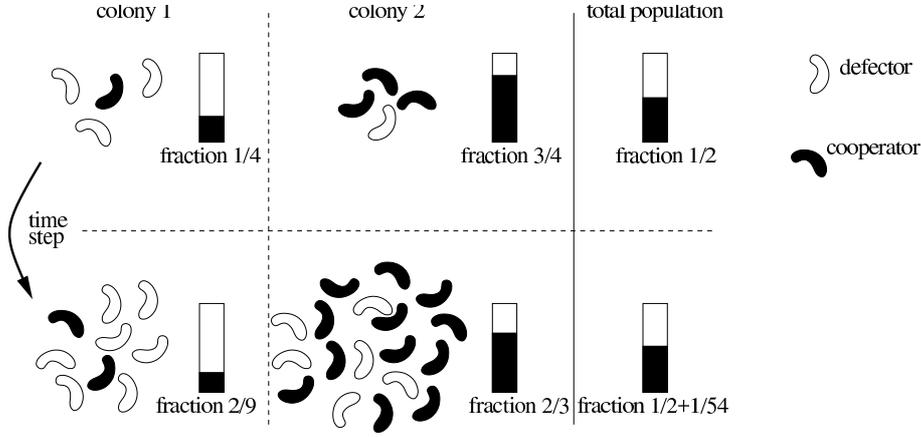}
\end{center}
\caption{The fraction of the cooperator becomes smaller in each colony, but 
larger in the total population (Simpsons paradox, see text). The reason is, that the colony with only 
few cooperators grows much slower than that with many cooperators.  
}\label{simpson}
\end{figure}

\medskip
Though in each colony the fraction of cooperators decreases, the colonies with many cooperators grow much faster than those with only few cooperators. Therefore, if $r$ is large, cooperation may spread (see figure~\ref{simpson}). This explanation is a  variation of the Simpsons paradox~\cite{Hauert2002,Chater2008,Chuang2009}\\

The next proposition is basically discussed in~\cite{Allen2016}, using 
a slightly different notation. To be self-contained, 
we reproduce the result using our notation.

\begin{propx} \label{qsInD} Let the resident be D, and the invader be QS. 
(a) $F'(0)$ is maximized by the choice
$\ol Q_I(y) = \theta(y)$, where $\theta(y)=1$ for $y\geq y^*$, 
and  $\theta(y)=0$ for $y<y^*$, where
$$y^*=p + (1-p)\frac c b.$$
(b) If $\ol Q_I(y) = \theta(y)$, $E_g(y\theta(y))>0$ and 
$$ c< b,$$ 
then  QS can invade in D.
\end{propx}
{\bf Proof: } (a) We have $\ol Q_R(x)=0$, and $\ol Q_I(y)\in[0,1]$. Then, 
\begin{eqnarray*}
 F'(0) &=& 
(1-p) E_g\left[y\,\left(-c\, \ol Q_I(y)+  b\,\ol Q_I(y)\,y      \right)\right]
-
p E_g\left[(1-y)\,b\,\ol Q_I(y)\,y \right]\\
&=&
E_g\bigg[
y\,\ol Q_I(y)\,\bigg(b\, (y-p)-c\,(1-p)
\bigg)
\bigg].
\end{eqnarray*}
As $y$ is non-negative, the expression is maximized if $\ol Q_I(y)=1$ for $b\, (y-p)-c\,(1-p)>0$, and $Q(y)=0$ if $b\, (y-p)-c\,(1-p)<0$. If $b\, (y-p)-c\,(1-p)=0$, the choice of $Q(y)$ does not matter. Now, $b\, (y-p)-c\,(1-p)>0$ is equivalent with $y>y^*$. (b) Thus, if $E_g(y\theta(y))>0$, then $F'(0)>0$.
\par\qed\par\medskip

Note that the optimization problem to maximize $E_g\bigg[
y\,\ol Q_I(y)\,\bigg(b\, (y-p)-c\,(1-p)\bigg)\bigg]$ 
resembles the optimization problem discussed in~\cite{Heilmann2015a}.  
We obtain the ``classical'' behavior expected of QS-bacteria: 
they are not cooperative if their population density is below a certain threshold, and jump to full cooperation if the quorum is reached. 

\begin{figure}[htb]
\begin{center}
\includegraphics[width=12cm]{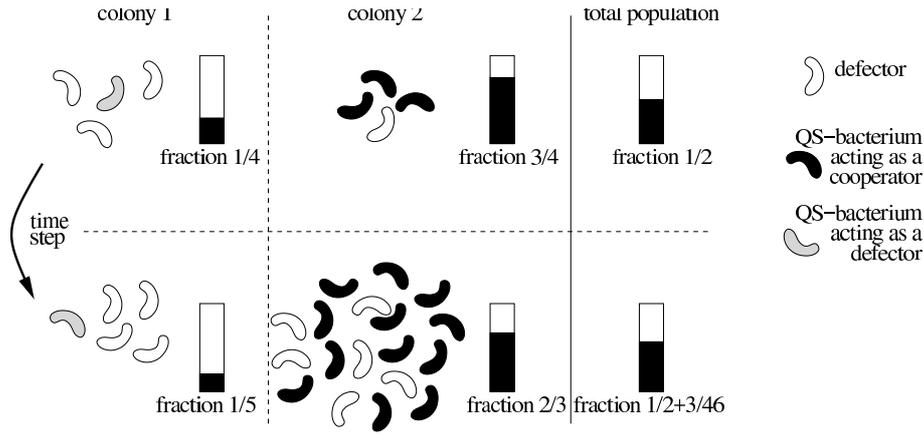}
\end{center}
\caption{The figure parallels figure~\ref{simpson}. Since the QS-individuals act as cooperators in colony 2, this colony evolves as colony 2 in figure~\ref{simpson}. In colony~1, the QS-individuals behave as defectors, such that (in the given example) almost no individual reproduces here. 
Consequently, after the time step, the ratio of QS individuals in the total population is larger than that of constitutive cooperator in figure~\ref{simpson}. The bars indicate the total fraction of QS individuals in the corresponding population (acting as cooperators or as defectors). 
}\label{simpsonQS}
\end{figure}

\begin{propx}\label{qsInC}  Let the resident be C, and the invader be QS. 
Assume that 
$\ol Q_I(y) = \theta(y)$, where $\theta(y)=1$ for $y\geq y^*$, and  $\theta(y)=0$ for $y<y^*$, where
$$y^*=p + (1-p)\frac c b.$$
If $c<b$ and $E(y(1-\theta(y)))>0$, then  QS can invade in C.
\end{propx}
{\bf Proof: }
We have $\ol Q_R(x)=1$, and $\ol Q_I(y)=\theta(y)$. Then, 
\begin{eqnarray*}
 F'(0) 
&=& 
(1-p) E_g\left[y\,\left(-c\, \theta(y)+  b\,(\, x+\theta(y)\,y)      \right)\right]
-
p E_g\left[(1-y)\,\left(-c\, +  b\,(\, x+\theta(y)\,y)      \right)\right]\\
&=& 
(1-p) E_g\left[y\,\left(-c\,+  b\,  + c(1- \theta(y))-by(1-\theta(y))    \right)\right]\\
&&-\quad
p E_g\left[(1-y)\,\left(-c\, +  b\, -by(1-\theta(y))   \right)\right]\\
&=& E_g\bigg[y\, (1-\theta(y))\bigg(c(1-p)-b y+b p\bigg)\bigg].
\end{eqnarray*}
As $1-\theta(y)=0$ for $y>y^*$ and $1-\theta(y)>0$ for $y<y^*$, and 
moreover, $c(1-p)-b y+b p>0$ for $y<y^*$, we have that $F'(0)\geq 0$. If, additionally, $E(y(1-\theta(y)))>0$, then $F'(0)>0$.
\par\qed\par\medskip

Naively, one would expect that QS individuals are always able to invade constitutive cooperators. However, if a QS individual fully cooperates, it has no advantage over cooperators (see figure~\ref{simpsonQS}). Stochasticity leads to the appearance of some colonies that start with only few QS bacteria. In these colonies, QS individuals will defect (at least initially), and have therefore an advantage over the cooperators. Using this mechanism -- the Simpson paradox  --   QS individuals can successfully invade a C population.\par\medskip

\begin{rem}
As we are interested in an invasion analysis, $E_g(I_0)=p$ is small.
Hence, in a good approximation, $y^* = c/b$ in proposition~\ref{qsInD} and proposition~\ref{qsInC}. At the end of the day, 
QS individuals do not cooperate until cooperation pays for each individual 
in the QS 
subpopulation: if $y>y^*$, then  $1-c\theta(y)+y\,b\theta(y)
> 1-c+b c/b = 1$. That is the case the fitness for the cooperating phenotype is larger than that for the 
defecting phenotype.   QS individuals do not cooperate 
in the altruistic sense, but they 
only produce PG if it pays for each single QS-individual. 
This observation is in line with the interpretation of quorum sensing 
as efficiency sensing~\cite{Hense2007}. The fact that other QS-bacteria or other 
bacterial species benefit from the cooperative phenotype is only a side effect. 
An individual maximizes its own fitness only.
\end{rem}

\subsection{Multilevel analysis for heterogeneously responding QS individuals}
\label{InvasionHetero}
We extend the setting above. First of all, in general, the impact of PG is nonlinear. As e.g.\ discussed in~\cite{Heilmann2015a}, it is important to take into account this nonlinear dependence. We use a function $B(.)$ to describe a non-linear beneficial effect of PG on the fitness of an individual.  Similarly, we assume that the costs do not depend linearly on the cooperativeness. We model the costs by a nonlinear function $C(.)$. 
Later we will see that concave and convex functions $C(.)$ are of special interest. 
We generalize our model also in a second point:  Each single QS-individual is allowed to choose its own 
degree of cooperativeness. In the last section, we had one single function $\ol Q_I(y)$ 
that indicated the degree of cooperativeness for all invading QS individuals in presence of a fraction $y$ of QS-individuals. This time, 
we have a vector $(Q_1(y),\ldots Q_\nu(y))^T\in[0,1]^\nu$ 
(if the number of QS-individuals is $\nu$), where $Q_i(y)$ represents the response of the $i$'th QS individual. 
We analyse the invasion of a defecting resident population by QS-bacteria. 
Let the colony equal $N$. Then, the fitness of a QS-individual with cooperativeness $Q$
 is given by 
$$ f_{QS}(Q; Q_1,\ldots, Q_\nu) = 1 - C(Q) 
+ \frac 1 N B\left(\sum_{i=1}^\nu Q_i\right)$$
while that for the defector is given by
$$ f_{D}(Q_1,\ldots, Q_\nu) = 1 
+ \frac 1 N B\left(\sum_{i=1}^\nu Q_i\right).$$
The degree of cooperation $Q$ is a central ingredient of the model. Costs and benefit are allowed to be non-linear functions on $\sum_{i=1}^\nu Q_i$ and $Q_i$, respectively. This general setting might introduce some arbitrary aspect. If we find $\tilde B$, $\tilde C$ and a monotonous function $g$, s.t.
$$ B\left(\sum_{i=1}^\nu Q_i\right) = \tilde B\left(\sum_{i=1}^\nu \tilde Q_i\right),\quad 
C(Q_i) = \tilde C(\tilde Q_i)$$
where $\tilde Q_i=g(Q_i)$, the interpretation of $Q_i$ as the normalized PG production rate might be difficult.  The next lemma shows that, under reasonable assumptions, only trivial rescaling of $Q_i$ is possible. We can choose the units used to measure $Q_i$ (namely the amount of PG contributed), but not more. 
\begin{lem} Assume that $B'>0$. If there is a function 
$\tilde B:\R_+\rightarrow\R_+$, $\tilde B'(x)>0$ and 
$\tilde B(0)=0$, and a function $g:\R_+\rightarrow\R_+$ 
rescaling $Q$, such that 
$$\forall \nu\in\N, (Q_1,\ldots,Q_\nu)\in[0,1]^\nu:\,\,\, B\left(\sum_{i=1}^\nu Q_i\right) =  \tilde B\left(\sum_{i=1}^\nu g(Q_i)\right)$$
then $g(x)=ax$ for some $a>0$, and $\tilde B(x)=B(x/a)$. 
\end{lem}
{\bf Proof: } Consider $\nu=3$, and $(Q_1,Q_2,Q_3)$, such that $\sum_{i=1}^3Q_i=z\in\R$. We keep $z$ fixed but vary the $Q_i$. Then, 
$$ B(z)=B(Q_1+Q_2+Q_3)=\tilde B(g(Q_1)+g(Q_2)+g(Q_3))$$
As $\tilde B'(x)>0$, $g(Q_1)+g(Q_2)+g(Q_3)$ is constant for all feasible triples $(Q_1,Q_2,Q_3)$. Replacing $Q_3$ by $z-Q_1-Q_2$ and taking the derivative w.r.t.\ $Q_1$, we obtain
$$ g'(Q_1)-g'(z-Q_1-Q_2)=0$$
As $Q_2$ can be chosen independently of $Q_1$ (as long as $Q_1+Q_2\leq z$), this equation can be only satisfied if $g'(x)$ is constant. Let $g'(x)=a$, then $g(x)=ax+b$. As $g:\R_+\rightarrow\R_+$, we have $a,b\geq 0$.\\
In order to determine $b$, we inspect $Q_1=Q_2=Q_3=0$. Therefore $z=0$, and 
$$ 0=B(0)=B(z)=\tilde B(3\,g(0))=\tilde B(3\,b).$$
Since $\tilde B(0)=0$, and $\tilde B'(x)>0$, we conclude $b=0$. The choice $a=0$ is not possible as otherwise $\tilde B\left(\sum_{i=1}^\nu g(Q_i)\right)$ does not depend on $Q_i$. Hence, $a>0$ and our original equation reduces to 
$$B\left(\sum_{i=1}^\nu Q_i\right) =  \tilde B\left(a\,\sum_{i=1}^\nu Q_i\right).$$
This equality implies $\tilde B(x)=B(x/a)$.
\par\qed\par\medskip 

We proceed with the investigation of the model. Thereto, we generalize the computation for the expression $F'(0)$ above: 
\begin{eqnarray*}
\frac d{dt}E_c(I_t|I_0=\nu)|_{t=0}
&=& 
\nu+\frac \nu N  B\left(\sum_{i=1}^\nu Q_i\right) - \sum_{i=1}^\nu C(Q_i),
\\
\frac d{dt}E_c(R_t|I_0=\nu)|_{t=0} 
&=&
(N-\nu)+\frac {N-\nu} N  B\left(\sum_{i=1}^\nu Q_i\right).
\end{eqnarray*}
Thus, with $p=E_g(I_0)/N$, 
\begin{eqnarray*}
F'(0)=\frac d{dt}\,\,\frac{E_g(I_t)}{E_g(I_t+R_t)}
&=& \sum_{\nu=0}^N P(I_0=\nu)\bigg[
\frac \nu N+\frac \nu {N^2}  B\left(\sum_{i=1}^\nu Q_i\right) 
-\frac 1 N  \sum_{u=1}^\nu C(Q_i)\\
&& \qquad\qquad\qquad-p\left(
1+\frac 1 {N}  b\left(\sum_{i=1}^\nu Q_i\right) 
-\frac 1 N  \sum_{u=1}^\nu C(Q_i)
\right)
   \bigg]\\
&=&
 \sum_{\nu=0}^N P(I_0=\nu)\bigg[
\left(\frac \nu N-p\right)\,\,\frac 1 {N}  B\left(\sum_{i=1}^\nu Q_i\right) 
-\frac {1-p} N  \sum_{u=1}^\nu C(Q_i)
   \bigg].
\end{eqnarray*}

\begin{corr}
	QS individuals can invade a  population of defectors if 
	\begin{eqnarray}
	F'(0)=  \sum_{\nu=0}^N P(I_0=\nu)\bigg[
	\left(\frac \nu N-p\right)\,\,\frac 1 {N}  B\left(\sum_{i=1}^\nu Q_i\right) 
	-\frac {1-p} N  \sum_{u=1}^\nu C(Q_i)
	\bigg]>0. 
	\end{eqnarray}
\end{corr}
Evolutionary forces will maximize $F'(0)$. 
In order to maximize $F'(0)$, the term in the brackets $[...]$  is to maximize for each $\nu$ separately. We are particularly interested in the invasion 
by QS individuals, that is, in $p\ll 1$. Therefore, we are allowed 
to neglect $p$ and are led to the following problem:

\begin{prob}\label{optiProblem}
Assume that there are $\nu$ QS-individuals and $N-\nu$ D-individuals. 
Find $Q\in[0,1]^\nu$, such that the function $\Lambda_\nu:[0,1]^\nu\rightarrow\R$, 
\begin{eqnarray}\label{optiProblemFunct}
\Lambda_\nu(Q) =  \frac \nu N\,\,  B\left(\sum_{i=1}^\nu Q_i\right) 
-  \sum_{u=1}^\nu C(Q_i)
\end{eqnarray}
is maximized.
\end{prob} 

\begin{rem}
If we inspect this optimization problem, we find that 
$$\nu+\Lambda_\nu(Q)
=
\sum_{i=1}^\nu \left[ 
1 +\frac 1 N\, B\left(\sum_{\ell=1}^\nu Q_\ell\right)-C(Q_i)
\right]
= \sum_{i=1}^\nu f_{QS}(Q_i; Q_1,\ldots,Q_\nu)
$$ 
is the total fitness of the QS subpopulation. If $\Lambda(.)$ is maximized, the total (and in this also the average) fitness of the QS subpopulation is maximized. This result is in line with Fisher's theorem of selection~\cite{Fisher1930,Hofbauer1998}. This theorem states that evolution favors strategies which maximize the average fitness of a population. Note here the difference between optimization of individual fitness and average population fitness. 
\end{rem}

Since $[0,1]^\nu$ is compact, $B$ and $C$ are continuous, we know that there is 
at least one solution to this optimization problem.
\begin{corr} There is $Q^*\in[0,1]^\nu$ such that $\forall Q\in[0,1]^\nu:\,\, \Lambda_\nu(Q)\leq\Lambda_\nu(Q^*)$.
\end{corr}

\begin{figure}[htb]
\begin{center}
\includegraphics[width=14cm]{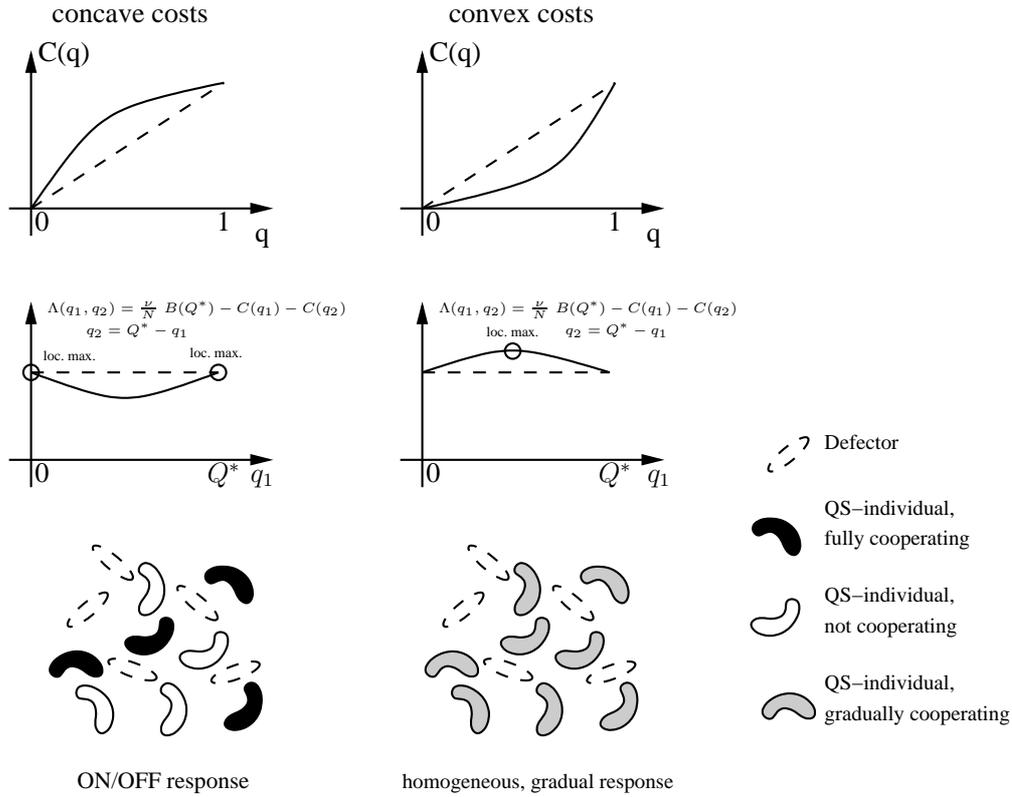}
\end{center}
\caption{Effect of concave (left) resp.\ convex (right) costs. First row: $C(q)$ over $q$. Second row: Initial rate $\Lambda_2$ for two cells and given $Q^*$, $\Lambda_2(q_1,q_2)=\frac \nu N B(Q^*)-C(q_1)-C(q_2)$ with $q_1+q_2=Q^*$, $q_1\in[0,Q^*]$.
Third row; left: we obtain a heterogeneous ON/OFF response; right: we obtain 
a gradual response.
}\label{convexConcave}
\end{figure}

There are, in particular, two cases, where (at least one) optimum 
$Q^*$ can be characterized: if $C$ is convex resp. concave (see figure~\ref{convexConcave}).
\begin{propx}\label{concavProp}
If $C(.)$ is concave then there is one optimal solution 
$Q^*$, where all components are in $\{0,1\}$ but one. 
If additionally $B(.)$ is convex, then all components are in $\{0,1\}$.
\end{propx}
Before we show this proposition, we recall some well-known terms: a set in $\R^n$ 
is convex, if the connecting line is always contained in the set. 
A convex simplex is the convex combination of a finite number of vertices, also
called extremal points. 
A function 
$f$ is convex, 
if the connecting line between two points on the graph of $f$ always 
lies above or on the graph, $f(\tau x+(1-\tau) y)\leq \tau f(x)+(1-\tau) f(y)$ 
for $\tau\in[0,1]$.  
A function $g$ is concave, if the converse is true, 
$g(\tau x+(1-\tau) y)\geq \tau g(x)+(1-\tau) g(y)$. If $f$ is convex, then $-f$ is concave and {\it vice versa}. 
If $f$ is twice differentiable and $x\in\R$, $f$ is convex (concave) if $f''\geq 0$ ($f''\leq 0$). If $x$ is element of a convex set in $\R^n$, $f$ is convex (concave) if the Hessian matrix is positive (negative) semidefinite. 
A convex function 
over a convex simplex assumes its maximum in a vertex (extremal point) 
of the simplex. A vertex can be characterized by the fact that it cannot 
be represented by the convex combination of two other points in the simplex.\\
The extremal points of $[0,1]^\nu$ consists of all vectors with entries in $\{0,1\}$. As the last point of this interlude, we state a well-known lemma about the 
extremal points of a certain convex set. Let $\e=(1,\ldots,1)^T$. For 
a simplex set ${\cal S}$, let $\Sigma({\cal S})$ denote the vertices 
of a simplex ${\cal S}$. 
\begin{lem}\label{lemmaExtrem} Let $\theta\in\R$, and 
${\cal S}=\{x\in[0,1]^n\,|\,\e^T x=\theta\}\not=\emptyset$.
If $\hat x\in\Sigma({\cal S})$ is an extremal point, then at least $n-1$  
entries of $x$ 
are in $\{0,1\}$. 
\end{lem}
{\bf Proof: } Assume $\hat x\in\Sigma({\cal S})$, and assume furthermore 
that $\hat x$ has at least two components $i_0$, $i_1$ in the open interval $(0,1)$ (where
$i_0\not = i_1$). Let $z_{i_0}=-1$, $z_{i_1}=1$, and $z_{i'}=0$ for 
all $i\not = i_0, i_1$. 
Then, $\hat x \pm \varepsilon z\in{\cal S}$ for $\varepsilon>0$ small enough. 
Thus, $0.5 ( (\hat x + \varepsilon z)+ (\hat x - \varepsilon z)) = \hat x$, in contradiction to the assumption that $\hat x\in\Sigma(\cal S)$.
\par\qed\par\medskip

{\bf Proof: } [of proposition~\ref{concavProp}] 
We know there is an optimum $Q^*$. 
Let 
$\overline Q^*(\nu) :=  \sum_{i=1}^\nu Q^*_i$. Then, $Q^*$ maximizes 
$\Lambda_\nu$ in particular in the simples 
${\cal S} = \{Q\in[0,1]^\nu\,|\, \sum_{i=1}^\nu Q_i=\overline Q^*(\nu)\}$. Note $Q^*\in{\cal S}$. 
For $Q\in{\cal S}$, we have 
$$\Lambda_\nu(Q) =  \frac \nu N\,\,  B\left(\overline Q^*(\nu)\right) 
-  \sum_{u=1}^\nu C(Q_i).$$
The first term is constant on ${\cal S}$. 
As $C(.)$ is concave,  $\Lambda_\nu(.)$ is convex on ${\cal S}$. 
There is an extremal point of ${\cal S}$ where  
$\Lambda_\nu(.)$ assumes its maximum on ${\cal S}$.  With lemma~\ref{lemmaExtrem} the first 
result follows.\\
If additionally $B$ is convex, then $\Lambda_\nu(.)$ is convex on $[0,1]^\nu$, and hence there is an extremal point of $[0,1]^\nu$ that maximizes $\Lambda_\nu(.)$. 
\par\qed\par\medskip

\begin{propx}\label{convexProp}
If $C(.)$ is convex, then there is one optimal solution 
$Q^*$, where all components are equal.
\end{propx}
{\bf Proof: } We use the nomenclature of proposition~\ref{concavProp}. 
On ${\cal S}$, $\Lambda_\nu(.)$ is concave. 
Let $q^*=\overline Q^*(\nu)/\nu$, and $\tilde Q^*=(q^*,\ldots, q^*)$. 
Then for $Q\in{\cal S}$ (that is, $\sum_{i=1}^\nu Q_i=\overline{Q}^*(\nu)$) 
\begin{eqnarray*}
 \sum_{i=1}^\nu C(Q_i) 
= \nu\, \sum_{i=1}^\nu \frac 1 \nu C(Q_i)
\geq  \nu\, C\left(\sum_{i=1}^\nu \frac 1 \nu Q_i\right)
=  \nu\, C(\overline{Q}^*(\nu)/\nu)
= \nu C(q^*).
\end{eqnarray*}
Thus, for $Q\in {\cal S}$, we have $\Lambda_\nu(Q)\leq \Lambda_\nu(\tilde Q^*)$. Since $\Lambda_\nu$ assumes its maximum in ${\cal S}$, the result follows.
\par\qed\par\medskip

\begin{rem}
If $B(.)$ and $C(.)$ are linear, propositions~\ref{concavProp} and~\ref{convexProp} are true at the same time. That is, either 
$Q^*=(0,\ldots,0)$, or $Q^*=(1,\ldots,1)$. Under non-favorable conditions, 
no QS-individual cooperates, while under favorable conditions all cooperate. 
We recover the result from section~\ref{manyColonies}.
\end{rem}
The next, natural question addresses the dependency 
of the  total amount  of cooperativeness in the optimal
strategy on the number (fraction) of cooperators in a colony. 
Recall that 
$$ \overline Q^*(\nu) :=  \sum_{i=1}^\nu Q^\ast_i$$
is the total amount of cooperativeness among $\nu$ QS-individuals. 
If $C(.)$ is convex, and $B(.)$ concave
(that is, the all-or-nothing strategy is optimal), 
$\overline Q^*(\nu)$ is precisely the number of cooperating cells.

\begin{propx} \label{overleinQ1}
If $C(.)$ is convex, $B(.)$ concave, and $Q_i\in\{0,1\}$, then
$\overline Q^*(\nu)$ is non-decreasing in $\nu$.
\end{propx}
{\bf Proof: }
Since $C$ is convex, and $B$ concave, $\Lambda_\nu$ is concave, and 
hence for the maximizing solution, $Q_i\in\{0,1\}$. All cells are either ON or OFF. \\ 
For symmetry reasons, only the number of cooperating cells matter. 
Given $\nu$, we may switch on one cell after the other. If $i$ cells are 
ON, then the value of $\Lambda_\nu$ is given by 
$\lambda_\nu(i) := \nu B(i)/N-i\, C(1)$. The additional net benefit (i.e., benefit minus costs) of the $i$'th cell thus reads
$$\Delta\lambda_\nu(i) = \lambda_\nu(i)-\lambda_\nu(i-1)
= \nu (B(i)-B(i-1))/N-C(1).$$
Since $B(.)$ is increasing, $B(i)-B(i-1)\geq 0$. As $B(.)$ is concave, 
the increase in the benefit decreases in $i$, 
$$ B(i)-B(i-1) \geq B(i+1)-B(i).$$
Thus, the number of active cells $i^\ast$ is either given by $i^\ast=\nu$, 
or by the condition $\Delta\lambda_\nu(i^\ast)\geq 0 >\Delta\lambda_\nu(i^\ast+1)$.
We have 
$$\Delta\lambda_{\nu+1}(i) =  (\nu+1) (B(i)-B(i-1))/N-C(1)\geq 
 \nu (B(i)-B(i-1))/N-C(1)= \Delta\lambda_{\nu}(i). $$ 
That is, if the net benefit of the $i$'th individual is positive in the population 
of $\nu$ QS-individuals, it is also positive in the population of $\nu+1$ 
Qs individuals. Thus, $\overline Q^*(\nu)$ is non-decreasing.
\par\qed\par\medskip

\begin{propx}  \label{overleinQ2}
If $C(.)$ is concave, and $B(.)$ convex, then
$\overline Q^*(\nu)$ is non-decreasing in~$\nu$.
\end{propx}
{\bf Proof: }
The basic argument parallels that from proposition~\ref{overleinQ1}. This time, 
we know that all cells do the very same, $Q^*_i=q^*$, according to the optimal strategy. 
We define $\lambda_\nu(q) = \nu(B(\nu\,q)-C(q))$. 
Then, $\lambda_\nu(q)$ is convex. As $q^*$ maximizes this function, we 
take $q^*=0$ if  $\lambda_\nu(q)$ is decreasing on $[0,1]$ (case 1); if it is increasing 
on $[0,1]$, we take $q^*=1$ (case 2); in case~3, we know that $\lambda_\nu(q)$
is an unimodal function with an internal maximum; $q^*\in(0,1)$ is located at this maximum.\\ 
In case 1, $\overline Q^*(\nu)=0$ and hence $\overline Q^*(\nu+1)\geq \overline Q^*(\nu)$.\\
For case 2 and 3, compare the derivatives of $\lambda_\nu(q)$ and $\lambda_{\nu+1}(q)$. Thereto note that $B'(.)>0$. As $B(.)$ is convex, $B'(.)$ is increasing. If $\frac d {dq} \lambda_\nu(q)\geq 0$, then 
\begin{eqnarray*}
0&\leq& \frac d {dq} \lambda_\nu(q) = 
 \nu(B'(\nu\,q)\,\nu-C'(q)) \leq 
(\nu+1)\,(B'(\nu\,q)\,\nu-C'(q))\\
&\leq&  (\nu+1)\,(B'((\nu+1)\,q)\,\nu-C'(q)) 
\leq  (\nu+1)\,(B'((\nu+1)\,q)\,(\nu+1)-C'(q))\\
& = & \frac d {dq} \lambda_{\nu+1}(q).\end{eqnarray*}
That is, if the net benefit grows for a given $q$ in a population of size $\nu$, it also does so in a population of size $\nu+1$. Hence, the optimal value $q^*$ is non-decreasing 
in $\nu$, and thus
$$ \overline Q^*(\nu+1)\geq \frac{\nu+1}{\nu}\, \overline Q^*(\nu)\geq \, \overline Q^*(\nu).$$
\par\qed\par\medskip

\begin{rem} The maximization problem~\ref{optiProblem} not only applies to a multilevel evolution analysis. If the average fitness of QS individuals within one single, spatially well mixed population is to be maximized, we are faced with the very same optimization problem. That is, also in the case of a single population, we find a heterogeneous/homogeneous response to be optimal, in dependence on the shape of the costs. However, QS individuals still cannot invade a well-mixed population of defectors, as the benefit is socialized while the costs are only carried by the QS trait. The maximization of the average fitness does not necessarily yield an evolutionary stable state (ESS). 
\end{rem}

Above we did focus on single bacteria. Now we switch the point of view, and address the total population. It is of particular interest to understand if the response of the population, which can be measured by the total PG production rate $\sum_{i=1}\nu Q_i$, is graded, or all-or-nothing. Basically we want to know what circumstances favor either a sudden, discontinuous increase in the total PG production rate if we increase $\nu$, or a continuous increase. In the strict sense, this question is meaningless as $\nu\in\{0,\ldots N\}$ is discrete. In order to identify the possibility for a response depending continuously (in a graded way) on $\nu$, we interpolate the optimization problem and allow for  $\nu\in[0,N]$.\\
Case 1: Convex costs, homogeneous response on the population level, $Q_i\equiv Q$ for $i\in\{0,\ldots,\nu\}$. 
Hence, $\sum_{i=1}^\nu Q_i = \nu\, Q$ and $\sum_{i=1}^\nu C(Q_i)=\nu\, C(Q)$. Therefore, eqn.~(\ref{optiProblemFunct}) reads
\begin{eqnarray*}
\Lambda_\nu(Q) =  \nu\,\bigg\{\frac 1 N\,\,  B(\nu\, Q) - C(Q)\bigg\}
\end{eqnarray*}
Since we aim to find $Q\in[0,1]$ that maximizes this expression, we can maximize as well ($\nu\in[0,N]$ fixed)
\begin{eqnarray}\label{optiProblemA}
	\tilde \Lambda_\nu(Q) = \frac 1 N\,\,  B(\nu\, Q) - C(Q).
\end{eqnarray}
Case 2: Concave costs, ON/OFF response on the individual level, $Q_i\in\{0,1\}$. According to proposition~\ref{concavProp}, there might be one single bacterium with $0<Q_i<1$; if $\nu$ is large, we can neglect this single bacterium and assume $Q_i\in\{0,1\}$. Let $\eta$ denote the fraction of QS bacteria in the ON state, that is, the number of ON bacteria are $\eta\,\nu$. Then, $\sum_{i=1}\nu Q_i=\eta\nu$, and $\sum_{i=1}^\nu C(Q_i)=\nu\, \eta C(1)$. 
In the second case, eqn.~(\ref{optiProblemFunct}) can be written as (note that $Q=(Q_1,\ldots,Q_\nu)$ is characterized by $\eta$)
\begin{eqnarray*}
	\Lambda_\nu(Q) =  \nu\,\bigg\{\frac 1 N\,\,  B(\nu\, \eta) - \eta\,C(1)\bigg\}
\end{eqnarray*}
where we aim to find $\eta\in[0,1]$ maximizing $\Lambda_\nu$. As before, we may consider as well 
the maximization problem ($\nu\in[0,N]$ fixed)
\begin{eqnarray}\label{optiProblemB}
\hat \Lambda_\nu(\eta) = \frac 1 N\,\,  B(\nu\, \eta) - \eta\,C(1).
\end{eqnarray}
If we compare the two optimization problems for case~1 (eqn.~(\ref{optiProblemA})) and 
case~2 (eqn.~(\ref{optiProblemB})), we find that they look very similar. $Q\in[0,1]$ is the normalized 
PG production, identical in all cells, while $\eta\in[0,1]$ refers to the fraction of activated cells. In that, $Q$ and $\eta$ characterize the activation level of the population. Experimental results (see discussion below) indicate that the costs depend in first order linearly on the PG produced, and convexity/concavity of the cost function is due to a higher order perturbation. In this case,  $C(Q)\approx Q\,C(1)$. That is, (\ref{optiProblemA}) and (\ref{optiProblemB}) have a similar solution. If the optimization problem is stable (in the sense that a small perturbation of the functional that is to optimize only leads to a small perturbation of the solution), then $Q\approx \eta$. In both cases, the optimal PG production is a certain fraction of the maximal PG production possible. In convex cost, all cells contribute equally to the PG production, while in concave costs, a fraction of QS individuals contribute at maximal rate, while the others do not contribute at all. \par\medskip 

\begin{figure}[htb]
\begin{center}
	\includegraphics[width=12cm]{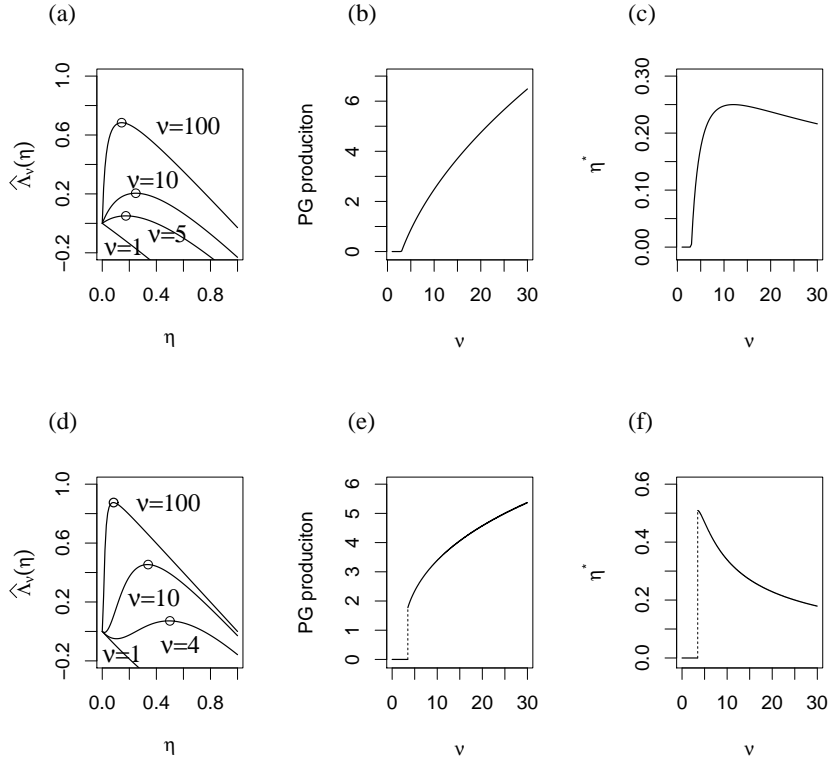}
\end{center}
\caption{(a), (d) Net benefit $\hat\Lambda_\nu(\eta)=B(\nu\eta)/N-\eta C(1)$ over $\eta$, where the maxima for the different curves ($\nu>1$) are indicates by circles; (b), (e) toal PG production for the optimal $\eta^\ast$ over $\nu$; (c), (f) optimal $\eta^\ast$ over $\nu$. 
Parameters used: 
$a = 100$, 
$b = 3$, $N=100$. First row: $n=1$, second row: $n=2$.
 }\label{optiFig}
\end{figure}

A typical choice for $B(.)$ is a Hill function, 
$$B(x) = \frac{a x^n}{b + x^n}.$$
The saturation of the benefit models spatial effects, while $n>1$ models synergistic effects. 
As depicted in figure~\ref{optiFig}, we expect a graded response for $n=1$, while a jump-like response is possible for $n=2$. While the total PG production is non-decreasing in $\nu$ (in accordance to the proposition above), the fraction of activated cells does not monotonously depend on $\nu$. The unimodal shape is a consequence of the saturating benefit. These findings are in line with similar investigations in~\cite{Heilmann2015a}.
\par\bigskip 

From these considerations we obtain the following picture: The total PG production is a non-decreasing function on the number of QS-individuals in the population. If the number of QS individuals is increased, the total PG production may increase continuously in a gradual response, or the total PG production is switched on at a certain threshold. Note, that even if the population responds discontinuously, in many cases the population does not jump to the maximal PG production possible, but only jumps to an intermediate PG production rate (see figure~\ref{optiFig} (e), (f)). Which of the two cases is realizes mainly depends on the nonlinear shape of the benefit $B(\,.\,)$. In contrast, how single bacteria contribute to the total PG production mainly depends on the shape of the costs $C(\,.\,)$: For convex costs, all QS individuals contribute at the same degree, while for concave cost functions a fraction of bacteria contribute at maximal rate while the other do not contribute at all. Of course, these different responses at single-cell-level  are only possible if not all cells are in the OFF or in the ON state.

\subsection{The shape of costs}
The present article does not aim to model molecular mechanisms, but concentrates on evolution. However, as the shape of the cost function is central, we discuss some biochemical aspects more in detail that may contribute to a nonlinear shape of the cost function.\\
There is experimental evidence that in many cases the running costs for protein production is proportional to the rate at which the protein is formed~\cite{Kafri2016,Harrison2012a,Glick1995}. These results can be simply understood: The production of one single enzyme has given costs $c_e$, and this enzyme produces PG at rate $\gamma$ for a certain time $T$. The production of one unit PG is given by some constant $c_t$. The total costs caused by one enzyme over time $T$ is given by $c_t\, \gamma\,T+c_e$, and the amount of PG produced by $\gamma\,T$. The costs per PG is $c_e/(\gamma\,T)+c_t$.  In order to have more PG at time $T$, we need more enzymes. The resulting relation between the amount of PG and the total costs is strictly linear, 
$$C(\mbox{PG}) = \frac{c_e}{\gamma\, T}\, \mbox{PG}+c_t\, \mbox{PG}.$$
The second term, $c_t\,\mbox{PG}$, relates to the actual PG production costs, while the first term is due to the ``infrastructure'' (enzyme production).\\ 
In our thought model, we implicitly made two assumptions: (a) The costs per active enzyme is independent on the number of enzymes, and (b) all enzymes are active over the complete time interval $[0,T]$. Both assumptions are not necessarily given. In the following, we discuss these two assumptions based on toy models. Afterwards, we consider the effect of incompatible metabolic processes and of privileged share on the cost function.\par\medskip 

Costs per active enzyme can be non-linear: A famous PG under the control of QS is light production by {\it Vibrio fischeri}. Light production is due to luciferase,  a heterodimer that consists of two subunits, $\alpha$ and $\beta$. The subunits are coded by the genes {\it luxA} and {\it luxB}~\cite{clark1993}, which are located in the lux cassette of {\it Vibirio fischeri}. The heterodimer luciferase $[\alpha\beta]$ is the result of an equilibrium reaction~\cite{clark1993}, $\alpha+\beta \rightleftharpoons[\alpha\beta]$. 
We neglect that the situation is even more complex, as luciferase has several conformations~\cite{clark1993}. The equilibrium of the reaction is given by
$$ K=\frac{[\alpha\beta]}{\alpha\,\beta}$$
for some constant $K$. If we assume that a cell produces the same number $n$ of both units, then $n=\alpha+[\alpha\beta]=\beta+[\alpha\beta]$. The number of luciferase molecules reads 
$$ [\alpha\beta] = K\,\alpha\,\beta = K\,(n-[\alpha\beta])^2\qquad\Rightarrow\qquad 
n = [\alpha\beta] +\sqrt{\frac{[\alpha\beta]}{K}}.$$
Even if the enzyme itself is the PG (as in the excretion of exoenzymes), this formula indicates that the costs per unit PG are concave. However, we proceed to discuss a PG produced by the enzyme. 
For a time interval $[0,T]$, the number of PG units produced are $\mbox{PG}=\gamma\,T\,[\alpha\beta]$, and the total costs induced are $C=c_e\,n +c_t\,\gamma\,T\,[\alpha\beta]=c_e\,n+c_t\,\mbox{PG}$. Hence, 
$$ n = 
C(\mbox{PG}) = c_e\bigg\{ \frac 1 {T\gamma}\,\,\mbox{PG} +\sqrt{\frac{1}{\gamma\,T\,K}}\,\sqrt{\mbox{PG}}\bigg\}\, + \,\,c_t\,\mbox{PG}.$$
As the second derivative of $C(\mbox{PG})$ is positive, this function is concave. We have one term that expresses the direct costs for the PG produced, and in addition a non-linear term that characterizes the efficiency of the pathway in dependency of its activity. The costs for the ``infrastructure'' of the pathway becomes non-linear. \par\medskip

Timing of enzyme production can induce nonlinear costs: We again formulate a most simple model to investigate this effect. Let $e(t)$ denote the amount of operating enzymes within a cell, and $\mbox{pg}(t)$ the amount of PG produced by this cell until time $t$. This PG could be, for example, exopolysaccharides. If $\eta$ is the production rate of enzymes, while enzymes produce PG at rate $\gamma$, we have 
$$ \dot e = \eta,\qquad \dot{\mbox{pg}}=\gamma\,e.$$
The total costs at time $t$ are due to enzyme production ($c_e\, \eta$) and due to PG production ($\gamma\,e\,c_t$). Hence, for the metabolic costs $c(t)$ incurred in the time interval $[0,t]$, we have 
$$\dot c = \eta\,c_e+\gamma\,e\,c_t.$$
We start with $e(0)=\mbox{pg}(0)=c(0)=0$. Therefore,  $e(t)=\eta\, t$. At time $T$ we obtain the total amount of PG produced and the total costs $C$
$$ \mbox{PG} = \frac 1 2 \,\,\gamma\,\eta\, T^2,\qquad C=\eta\,c_e\,T+\frac 1 2 \,\,\gamma\,\eta\, c_t\,T^2.$$
Therewith, 
$$C(\mbox{PG}) = \frac{\sqrt{2\,\eta}\,c_e}{\sqrt{\gamma}}\sqrt{\mbox{PG}}\,+\,c_t\,\mbox{PG}.
$$
Again, we find a concave dependency of the costs on the amount of PG produced. And again, the costs consist of a term measuring the direct costs for the PG produced, and a second term related to the efficiency of an enzyme, that is, the average amount of PG produced per enzyme. Not the costs for PG production itself, but the costs related to the infrastructure are non-linear (concave) in the last two models. 
\par\medskip

Molecular mechanisms that may lead to a convex shape are completely different: The production effort per unit PG becomes increasingly higher with an increasing production rate. This is the case if e.g.\ waste products of the PG production accumulate and are detrimental for the cell. Ackermann~\cite{Ackermann2015} describes similar ideas to explain phenotypic heterogeneity if different, mutually exclusive tasks are to be performed by a bacterial population. \par\medskip 

A possibility to influence the shape of the (net) costs is privileged share: if part of the PG produced is kept private by the producing bacterium, then the net costs are given by the costs (in the strict sense) minus the benefit due to the part of the PG that is kept privately. In that, the shape of the net costs depends on both, the shape of the costs in the strict sense, and the shape of the benefit. It is possible to ``overwrite’’ convexity of the costs by privileged share.\par\medskip 

The practical importance of the effects discussed here are unclear. Their relevance needs to be determined by experimental approaches. 

\section{Discussion}

{\it The nonlinear characteristics of the benefit controls the PG production on the population level, while the nonlinear characteristics of the costs controls the PG production on the individual level.}  
If the number of QS individuals increases, the total PG production by the QS population is non-decreasing. The exact way how the total PG production changes with the number of QS individuals -- if there is a gradual increase, or a sudden jump from no PG production to a high PG production -- mainly depends on the shape of the benefit. In contrast, mainly the shape of the cost function determines how this total PG production rate is realized on the individual level. Given a number of QS individuals, a certain optimal degree of cooperativeness on the population level can be reached either when all cells contribute in the same way, or when some cells do not contribute at all, while others cooperate as much as possible. The optimal strategy depends on the shape of the cost function. If the PG production is discounted, such that for a cell the production of  two PG units is cheaper than that of two times the costs for one unit (concave costs), the population is better off if some individuals produce heavily and others not at all (heterogeneous strategy). If the situation were the other way around (convex costs), then we would predict a homogeneous response. \\
At a second glance, the finding that mainly the shape of the cost function is decisive for the individual response is intuitive: PGs are assumed to be available in the same amount to all individuals. All cells benefit in the same way, independently of the degree of cooperativeness a given cell chooses.  As the selection happens based on fitness differences, a benefit that improves the fitness of all cells uniformly is of minor importance for competition within a colony. As usual, this observation is true as long as the benefit and the costs contribute in an additive way to the fitness. We might speculate that in a first step a homogeneous response evolves, that in a second step, driven by evolutionary forces, is adapted to a slightly more efficient, heterogeneous response. However, in multilevel evolution, also the benefit difference between different groups is of importance. While costs play the larger role within groups, the net benefit is of importance in the competition between groups.\\

We speculate that convex costs are frequent, as e.g.\ enzymes necessary for PG represent an additional cost and have to be produced for PG production. Hence, we also expect that phenotypic heterogeneity in the PG production can be often found in nature.
\par\medskip

{\it QS populations with a high density of individuals may nevertheless not lead to a complete activation  of the population.} Experiments show that, even at high concentrations of QS individuals, we still find heterogeneity. Naively stated, we would expect that the total population of QS individuals is activated if the population density becomes large and very large. However, in~\cite{Anetzberger2009,Perez2010} it is reported that this is not the case for a population of {\it Vibrio harveyi} resp.\ {\it Vibrio fischeri}, through additional QS signals, are able to active also the non-induced part of the population. Also, this finding can be interpreted in the light of our theoretical study: Each increase of the cooperativeness comes not only with an increase of benefit but also with an increase of costs. The increase of the benefit is likely to be modeled by a saturation function. If the benefit approaches its plateau before all individuals are fully activated, it is advantageous for the population if not the maximal potential for PG production is exploited. In the case of concave costs (inducing a heterogeneous response), part of the population is fully induced, and another part not induced at all. The alternative is more expensive, i.e., that all individuals are homogeneously halted in their production, leading to the case of a homogeneous response (convex costs). The extra expense consists of the cost of production setup, e.g., enzymes necessary for PG production. In a homogeneous response, all individuals invest into the setup costs, and face the risk of not using them to their maximum potential.
\par\medskip

{\it Molecular mechanisms for heterogeneity.}
Positive feedback in many QS systems promoted an understanding that QS coordinates group behavior in an all-or-none fashion. This seems to contradict the idea of phenotypic heterogeneity~\cite{Hense2015}. However, recent publications revealed how all-or-none switching at the cellular level can lead to either all-or-none or graded responses at the population level~\cite{Fujimoto2013,Youk2014,Maire2015}. 
Stochastic differences in QS protein levels between cells, which might affect both signal production 
and/or perception, can be augmented by the positive feedback. If the coupling of cells via QS is very tight, 
then this heterogeneity can be overcome. However, the strength of coupling between cells, essentially 
the degree of self- or between-cell signaling, can be controlled. QS parameters such as signal production rate, 
secretion rate and receptor concentration determine to which degree the signal is shared or retained 
intracellularly~\cite{Fujimoto2013}. As a consequence, activation remains all-or-none at the cellular level, 
but can switch between all-or-none and graded (heterogeneous) at the population level. In fact, QS parameters 
have been shown to be under control of extracellular and intracellular conditions~\cite{Mellbye2013}. 
 \\
In the scenario described above, phenotypic heterogeneity emerges within the QS system itself. 
Alternatively, heterogeneity of QS controlled behavior can also be generated down-stream 
of QS signaling. The “Competence Stimulating Protein“ (CSP) in {\it Streptococcus} 
mutans evokes 
a response in all cells of a population by phosphorylation of ComE via activation of the histidine 
kinase ComD~\cite{Lemme2011}. Although the underlying mechanisms are not understood 
in detail, heterogeneities arising in the signaling cascade downstream of ComD finally result 
in a CSP-controlled induction of three phenotypic different subpopulations of competent, 
non-competent and autolytic cells. In {\it B. subtilis}, the comQXPA QS system regulates the rate 
of competence induction by controlling the competence master regulator ComK~\cite{Schultz2009}.
\par\medskip

{\it QS individuals are able to circumvent Hamilton's rule.} 
In a multilevel evolutionary setting, QS individuals  are able to invade residents which are 
defectors as well as constitutive cooperators. 
Constitutive cooperators can only invade defectors if Hamilton's rule $c<b \, r$ is given. 
QS individuals only require that $c<b$. For QS populations the relatedness $r$ does not play a role: Their advantage is that they can measure the relatedness~\cite{Schluter2016} in each single colony, and respond accordingly. We recovered and extended the result~\cite{Allen2016},  where competition experiments and the conceptual framework for the Price equation is used to show that QS individuals are able to fight cheaters by negative frequency-dependent selection. The assumption of a homogeneous response forces an even more restrictive interpretation: QS individuals with a homogeneous strategy do not really estimate the relatedness -- QS individuals are even more selfish. They  only start to cooperate, if cooperation pays for each single individual. QS with a homogeneous response can be viewed as a selfish 
strategy that seeks to maximize the fitness of single cells.
 \par\medskip

{\it Heterogeneous response is efficiency sensing in the light of Fisher's fundamental theorem of selection.} 
In the view of 
evolutionary pressure, a population  maximizes its average fitness. 
Therefore,  QS  is best interpreted as efficiency sensing~\cite{Hense2007}: the QS mechanism allows to 
estimate the effect of an action.  At this point, we need to refine the 
term ``effect''. Is this the effect  for the benefit of one single individual, 
or of the population? 
Often, one only considers the fitness of single individuals. However, fitness optimization 
of single individuals and fitness optimization of the population are only equivalent if there is no
interaction between the individuals. This is surely not the case in cooperation via PG. \\
If all individuals optimize their own fitness,  all individuals of an isogenic population exposed to the same
environment should behave in the very same manner. 
Experiments
as well as our modeling approach indicates that a heterogeneous phenotype 
can be favorable.  
Some QS-bacteria act as phenotypic ``cheaters'', benefiting from the 
PG produced by fellow bacteria, but not contributing to the PG production.   Quorum sensing 
can still be seen as efficiency sensing, but not maximizing the fitness of each single cell, but rather on the population level. According to 
Fisher's fundamental theorem of evolution, the mean fitness of a
population is maximized, not the fitness of single individuals~\cite{Hofbauer1998}. 
We are faced with a similar effect that stabilizes e.g.\ the altruistic behavior of persister cells. In 
the context of multiplayer game  theory, heterogeneous strategies appear if the benefit 
is chosen to be nonlinear, e.g.\ as a threshold function~\cite{Deng2011}.  
\par\medskip

{\it Acknowledgments} We thank Dieu-Thi Doan for useful discussions. 
This work was supported by the Deutsche Forschungsgemeinschaft  within the frame of the priority program SPP1617 (MU2339/2-2 to J.M. and Be 2121/6-1 to M.McI.), and by the National Science Foundation of the United States (MCB1616967 to M.S.).

\bibliographystyle{spmpsci} 
\bibliography{plasticityBib}

\end{document}